\documentclass[a4paper,fleqn,usenatbib]{mnras}

\usepackage{newtxtext,newtxmath}

\usepackage[T1]{fontenc}
\usepackage{ae,aecompl}

\usepackage{graphicx}	
\usepackage{amsmath}	
\usepackage{amssymb}	

\newcommand{\tone}{t_{\rm 1D}}
\newcommand{\tthr}{t_{\rm 3D}}

\title[Tidal double detonation]{Tidal double detonation: a new mechanism for a thermonuclear
  explosion of a white dwarf induced by a tidal disruption event}

\author[Ataru Tanikawa]{
Ataru Tanikawa,$^{1,2}$\thanks{E-mail: tanikawa@ea.c.u-tokyo.ac.jp}
\\
$^{1}$Department of Earth Science and Astronomy, College of
  Arts and Sciences, The University of Tokyo, 3-8-1 Komaba, Meguro-ku,
  Tokyo 153-8902, Japan\\
$^{2}$RIKEN Advanced Institute for Computational Science,
  7-1-26 Minatojima-minami-machi, Chuo-ku, Kobe, Hyogo 650-0047,
  Japan
}

\date{Accepted XXX. Received YYY; in original form ZZZ}

\pubyear{2017}

\begin{document}
\label{firstpage}
\pagerange{\pageref{firstpage}--\pageref{lastpage}}
\maketitle

\begin{abstract}

We suggest ``tidal double detonation'': a new mechanism for a
thermonuclear explosion of a white dwarf (WD) induced by a tidal
disruption event (TDE). Tidal detonation is also a WD explosion
induced by a TDE. In this case, helium (He) and carbon-oxygen (CO)
detonation waves incinerate He~WD and CO~WD, respectively. On the
other hand, for tidal double detonation, He detonation is first
excited in the He shell of a CO~WD, and drives CO detonation in the CO
core. We name this mechanism after the double detonation scenario in
the context of type Ia supernovae. In this paper, we show tidal double
detonation occurs in shallower encounter of a CO~WD with an
intermediate mass black hole (IMBH) than simple tidal detonation,
performing numerical simulations for CO~WDs with $0.60M_\odot$ with
and without a He shell. We expect tidal double detonation spreads
opportunity to WD~TDEs illuminating IMBHs.

\end{abstract}




\section{Introduction}
\label{sec:introduction}

A tidal disruption event (TDE) is a phenomenon in which a star is
tidally torn apart by a black hole (BH). There are many TDE candidates
in which main sequence (MS) stars are disrupted by massive black holes
(MBHs) \cite[e.g.][]{2015JHEAp...7..148K,2017ApJ...838..149A}. White
dwarfs (WDs) are also expected to experience TDEs. It is not by MBHs
but by intermediate mass black holes (IMBHs) that WDs are disrupted
\citep{1989A&A...209..103L}. WD~TDEs will produce not only bright
flares powered by accretion of WD debris onto IMBHs, but also
thermonuclear emissions of radioactive nuclei synthesized by tidal
detonation
\citep{1989A&A...209..103L,2004ApJ...610..368W,2008CoPhC.179..184R,2009ApJ...695..404R}.
WD~TDEs can be probes to explore IMBHs.

Tidal detonation happens in a WD~TDE as follows. A WD closely passes
by an IMBH, and is stretched by the tidal field of the IMBH in the
direction of the orbital plane (hereafter $x$-$y$ plane). On the other
hand, the WD is compressed in the direction perpendicular to the
$x$-$y$ plane (hereafter $z$-direction). This is because the WD size
become comparable to the separation between the WD and IMBH at the
pericenter. The compression of the WD has to accompany shock heating
for tidal detonation. In other words, adiabatic compression is
insufficient for tidal detonation. Although \cite{1989A&A...209..103L}
have suggested a helium (He) WD can be detonated only by adiabatic
compression, their He~WD model has relatively larger mass
($0.6M_\odot$) or higher density ($\sim 10^7$~g~cm$^{-3}$) than in
reality
\citep[e.g.][]{2017MNRAS.470.4473P}. \cite{2004ApJ...610..368W} have
reported a carbon-oxygen (CO) WD can experience tidal detonation by
adiabatic compression. However, they have not taken into account
elongation of the CO~WD by a tidal field of a BH, and have
overestimated the density of the CO~WD. Note that nuclear reactions
are more active under higher density environment.

In \cite{2017ApJ...839...81T} (Paper~I) and \cite{2017arXiv171105451T}
(Paper~II), we have investigated tidal detonation triggered by a shock
wave. We have found the following three facts. First, the shock wave
does not always result in a detonation wave. Second, the shock wave is
easier to excite a detonation wave under higher density environment
and in lighter nuclear compositions (in the order of He, CO, and
oxygen-neon-magnesium (ONeMg) compositions). Third, the shock wave
arises near the surface of a WD, when the ratio of the tidal
disruption radius to the pericenter distance ($\beta$) is not so
large. These facts imply a shock wave triggers tidal detonation more
easily from a CO~WD (ONeMg~WD) with a He shell than from a CO~WD
(ONeMg~WD) without a He shell. This tidal detonation proceeds as
follows. A shock wave generates He detonation in the He shell of a
WD. The He detonation invades into the CO core of the WD, and ignites
CO detonation in the CO core. We name this explosion mechanism ``tidal
double detonation'' after the double detonation scenario of type Ia
supernovae
\citep{1980tsup.work..164N,1982ApJ...257..780N,1980tsup.work...96W}.

The tidal double detonation is fairly a new scenario for a
thermonuclear explosion of a WD by a TDE. Tidal detonation considers
only He detonation in He~WD and only CO detonation in CO~WD
\cite[e.g.][]{2008CoPhC.179..184R,2009ApJ...695..404R}. There are
various double detonation scenarios for a thermonuclear explosion of a
WD
\citep{1980tsup.work..164N,1982ApJ...257..780N,1980tsup.work...96W,1990ApJ...354L..53L,2007ApJ...662L..95B,2010ApJ...709L..64G,2010ApJ...715..767S,2013ApJ...770L...8P}. However,
in these scenarios, He detonation is triggered by accretion of matter
from a companion star onto a WD.

In this paper, we show the tidal double detonation works better for a
thermonuclear explosion of a CO~WD than simple tidal detonation, if a
WD has a He shell whose mass fraction is $5$~\% of the WD. Note that a
CO~WD can have a He shell whose mass fraction is several \%
\citep{1985ApJS...58..661I,1993ApJ...418..343I,2000ApJ...544.1036S}.

The structure of this paper is as follows. We describe our method in
section~\ref{sec:method}. We present our results in
section~\ref{sec:results}. Finally, we make a conclusion in
section~\ref{sec:conclusion}.

\section{Method}
\label{sec:method}

We investigate tidal detonation and tidal double detonation in a
similar way to Paper~II. We follow overall evolution of a WD by
3-dimensional (3D) smoothed particle hydrodynamics (SPH)
simulation. We extract profiles of density and $z$-velocity in the
$z$-direction from a portion of the WD, and use the profiles for an
initial condition of 1-dimensional (1D) mesh simulation. We perform 1D
mesh simulation, and follow tidal detonation and tidal double
detonation. We combine 1D mesh simulation with 3D SPH simulation in
order to avoid spurious heating due to low space resolution
\citep[][Paper~I]{2008CoPhC.179..184R,2009ApJ...695..404R}, and in
order to resolve a shock wave near the surface of a WD.

Our 3D SPH code is the same as in Paper~II. We use FDPS
\citep{2016PASJ...68...54I} for parallelization of our 3D SPH code. We
adopt the Helmholtz equation of state (EoS)
\citep{2000ApJS..126..501T} which is calculated in the routine
developed by the Center for Astrophysical Thermonuclear Flashes at the
University of Chicago. We do not take into account nuclear reactions
in 3D SPH simulation. We follow the evolution of a CO~WD with
$0.6M_\odot$ by 3D SPH simulation. The composition is $50$\% carbon
and $50$\% oxygen in mass. The CO~WD has no He shell. The number of
SPH particles, $N$, is $100$ millions. We relax the configure of SPH
particles in the same way as in \cite{2015ApJ...807...40T}.  The IMBH
mass is $300M_\odot$. We adopt Newtonian potential for the IMBH
gravity. The orbit of the WD around the IMBH is parabolic. The ratio
of the tidal disruption distance to the pericenter distance ($\beta$)
is $5$. The IMBH does not irrupt into nor swallow the WD even if we
consider general relativistic effects for the IMBH gravity
\citep{2013MNRAS.433.1930T}. The WD passes the pericenter from
$\tthr=3.5$~s to $\tthr=4.5$~s, where $\tthr$ is the time from the
starting time of the 3D SPH simulation.

\begin{figure}
  \includegraphics[width=\columnwidth]{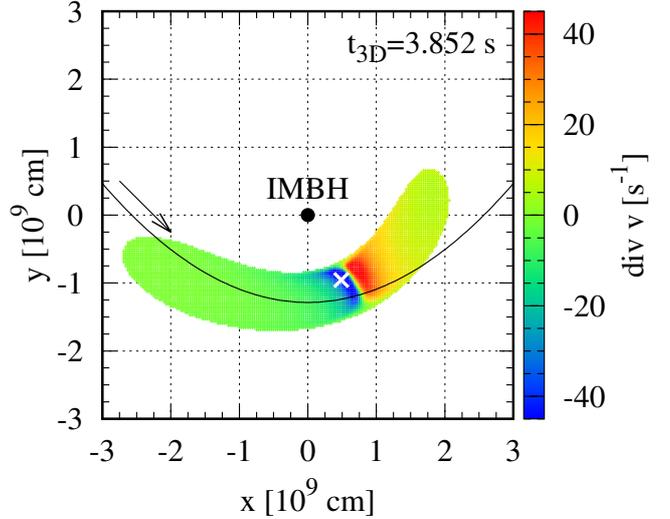}
  \caption{Divergence of velocity on the $x$-$y$ plane at the
    indicated time in the CO~WD. The IMBH is located at the coordinate
    origin. The solid curve shows the orbit of the WD on the
    assumption that the WD is a point mass, and the arrow indicates
    the traveling direction of the WD orbit. The white cross indicates
    the extracted portion.
\label{fig:init3d_divv}}
\end{figure}

We make 1D initial conditions, extracting density and $z$-velocity
profiles in the $z$-direction from a portion of the WD simulated by
the above 3D SPH method. We choose the portion indicated by the white
cross in Figure~\ref{fig:init3d_divv}. The portion will bounce back
immediately in the subsequent evolution. Owing to our choice of the
portion, we minimize the difference between 1D and 3D simulation
results which comes from 3D effects, such as a tidal field of the
IMBH. This is because we can follow the evolution of the portion for
as short a term as possible. Here, the evolution of the portion is
such that a pressure wave is generated at the bounce time, the
pressure wave steepens into a shock wave, and finally the shock wave
excites a detonation wave.

Using 3D SPH results of the CO~WD, we obtain density and $z$-velocity
profiles in the $z$-direction in the portion by applying SPH kernel
interpolation.  We set profiles of nuclear elements as follows. For a
CO~WD without a He shell, we assign only the CO composition to all the
regions for a 1D initial condition. For a CO~WD with a He shell, we
assign the CO composition to the inner region and the He composition
($100$\% He in mass) to the outer region. The mass fraction of the
outer region is $1$\% ($2$\%).  This is because we find the outermost
$1$\% ($2$\%) SPH particles in this portion form a He shell in the 3D
SPH simulation, assuming that the outermost $5$\% ($10$\%) SPH
particles in the initial CO~WD form a He shell in the 3D SPH
simulation. The mass fraction of He composition is much less than
$5$\% ($10$\%), since the geometry of the extracted portion is planar,
not spherical.

For 1D mesh simulation, we use the FLASH code
\citep{2000ApJS..131..273F}. We use uniform mesh. We adopt the
Helmholtz EoS and Aprox13 \citep{2000ApJS..129..377T} for our EoS and
nuclear reaction networks, respectively.
We suppress nuclear burning in shocked cells.
The calculation domain geometry is planar. The domain range is $0 \le
z/10^8\mbox{cm} \le 1.0$. The number of mesh is $6400$. The boundary
condition is the reflection condition at $z=0$~cm and the outflow
condition at $z/10^8\mbox{cm}=1.0$. The time from the starting time of
1D mesh simulation is indicated by $\tone$.

\section{Results}
\label{sec:results}

\begin{figure*}
  \includegraphics[width=2\columnwidth]{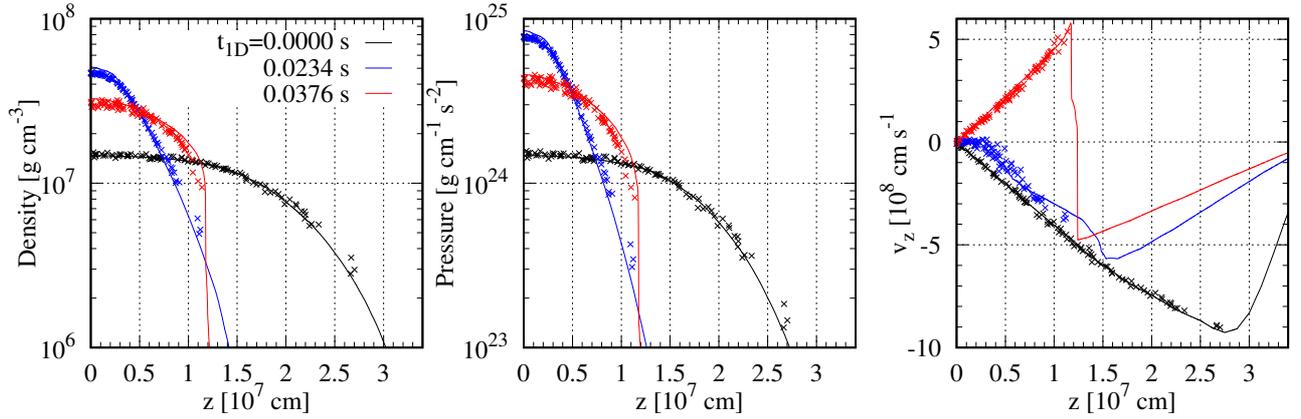}
  \caption{Time evolution of density, pressure, and $z$-velocity in 1D
    mesh simulation for a CO~WD without a He shell (solid curves) and
    in 3D SPH simulation (cross points). The cross points indicate
    physical quantities of SPH particles sampled randomly from the
    portion. At $\tone \sim 0.0234$~s, the portion bounces back, and
    at $\tone \sim 0.0376$~s, a shock wave
    appears. \label{fig:evolve1d_comp3d}}
\end{figure*}

We verify the evolution in the 1D mesh simulation mirrors the
evolution of the column in the 3D SPH simulation until a shock wave
appears. Figure~\ref{fig:evolve1d_comp3d} shows the time evolution of
density, pressure, and $z$-velocity in 1D mesh simulation for a CO~WD
without a He shell, and in 3D SPH simulation.
At $\tone = 0$~s, $z$-velocity gradient is positive at $z \gtrsim 3
\times 10^7$~cm, since we set the $z$-velocity to be zero outside the
WD.
At $\tone \sim 0.0234$~s, the portion bounces back, and at $\tone \sim
0.0376$~s, a shock wave appears. The evolution of physical quantities
in the 1D mesh simulation is in good agreement with the evolution of
physical quantities in the 3D SPH simulation. Despite that the 1D mesh
simulation ignores 3D effects, such as a tidal field, density and
pressure in the 1D mesh simulation are larger than those in the 3D SPH
simulation by $10$\% at $z=0$, and by $20$\% at the edge of the WD
when the shock wave appears at $\tone \sim 0.0376$~s. The $z$-velocity
in the 1D mesh simulation evolves in the same way as that in the 3D
SPH simulation. Later, we discuss whether the density overestimate
affects the initiation of a He detonation wave, or not.

\begin{figure*}
  \includegraphics[width=2\columnwidth]{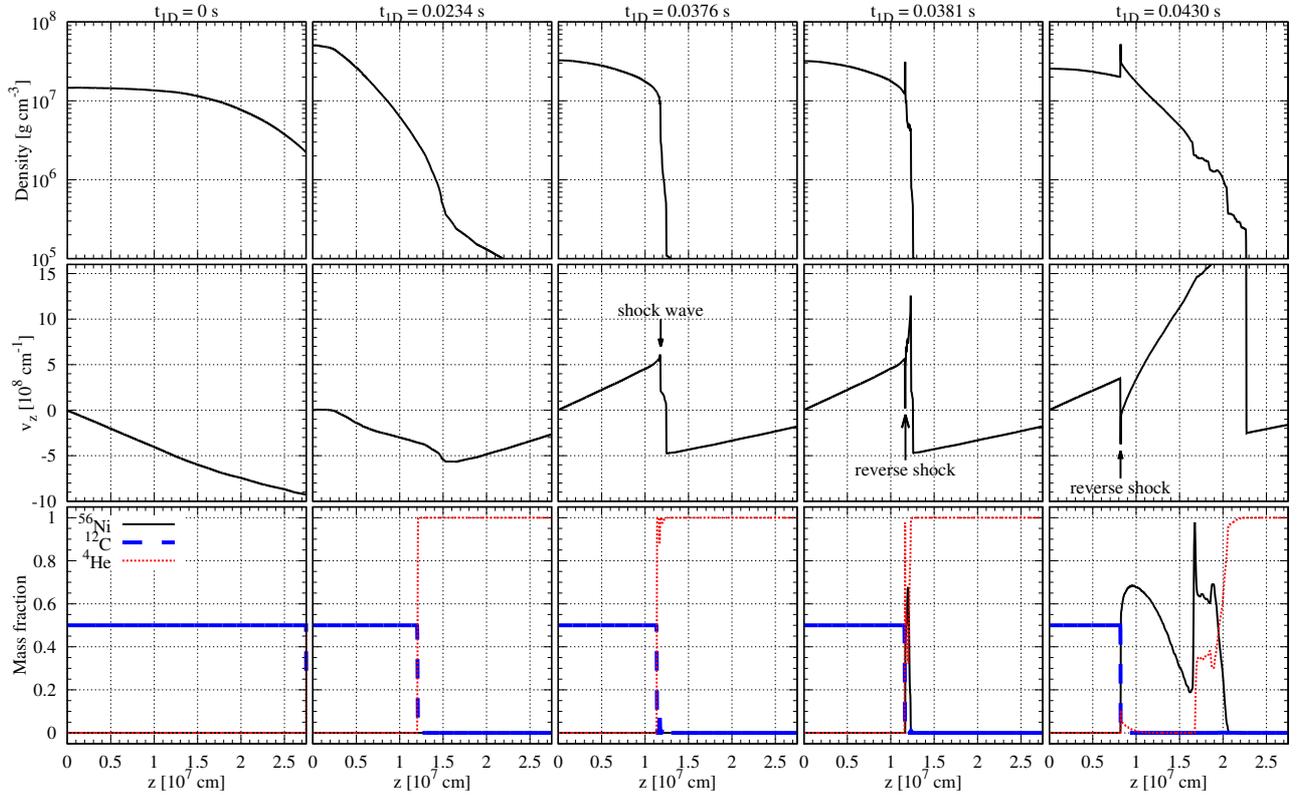}
  \caption{Time evolution of density, $z$-velocity, and nuclear
    element profiles in 1D mesh simulation for a CO~WD with a He
    shell. The mass fraction of the He shell is $2$\%. The initial
    condition is based on 3D SPH simulation of a CO~WD with
    $N=100$M. \label{fig:evolve1d_cohe}}
\end{figure*}

Figure~\ref{fig:evolve1d_cohe} shows the time evolution in 1D mesh
simulation for a CO~WD with a He shell. The mass fraction of the He
shell is $2$\%. In order to show tidal double detonation clearly, we
present the case where the mass fraction of the He shell is $2$\%.
From the beginning, the WD shrinks to the $x$-$y$ plane (or $z=0$). At
$\tone \sim 0.0234$~s, it bounces back, and a pressure wave arises. At
$\tone \sim 0.0376$~s, the pressure wave steepens into a shock wave
(see the middle panel at $\tone = 0.0376$~s), and the shock wave
triggers explosive nuclear reactions in the He shell (see the bottom
panel at $\tone = 0.0376$~s). The explosive nuclear reactions generate
a reverse shock wave (see the middle panel at $\tone = 0.0381$~s). The
reverse shock wave accompanies He detonation which consumes large
amounts of He materials (see the bottom panel at $\tone =
0.0381$~s). The He detonation invades into the CO core, and achieves
CO detonation in the CO core. In fact, the CO detonation burns out
large amounts of C materials (see the bottom panel at $\tone =
0.0430$~s). We also obtain the same results in the case where the mass
fraction of a He shell is $1$\%. Therefore, the CO~WD with a He shell
whose mass fraction is at most $5$\% succeeds tidal double detonation.

\begin{figure}
  \begin{center}
  \includegraphics[width=0.7\columnwidth]{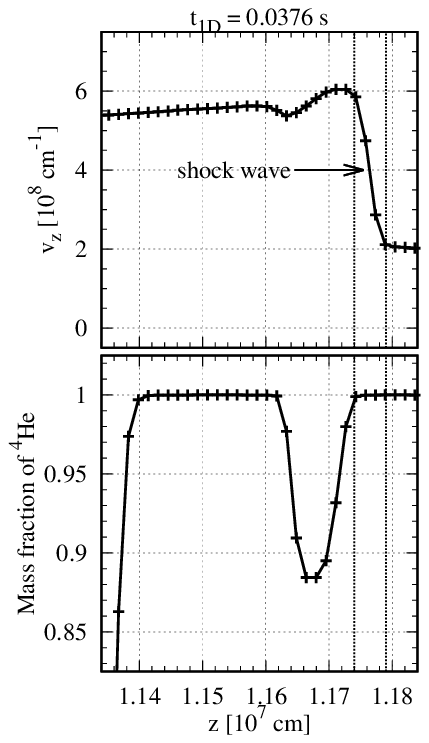}
  \end{center}
  \caption{Profiles of $z$-velocity and nuclear components at $\tone =
    0.0376$~s where the He detonation starts. Cross points indicate
    positions of cells in 1D mesh simulation. The shock wave indicated
    in Figure~\ref{fig:evolve1d_cohe} is between two vertical dotted
    lines. \label{fig:just_after_shock}}
\end{figure}

We investigate the initiation point of the He detonation in detail. In
Figure~\ref{fig:just_after_shock}, we zoom in on the position where
the He detonation starts. We can see the mass fraction of $^{4}$He at
$z \sim 1.166 \times 10^7$~cm is smaller than at its
surroundings. This is the initiation point of the He detonation. The
shock wave generating the He detonation is between the two vertical
dotted lines. Explosive nuclear reactions occur behind the shock
wave. This is because we suppress nuclear reaction networks in shocked
cells. The mass fraction of $^{4}$He is unity at the surroundings of
the initiation point of the He detonation. Therefore, the He
detonation starts under pure He environment. Even if the components of
the He shell and CO core are mixed, the mixing does not affect the
initiation of the He detonation.

\begin{figure*}
  \includegraphics[width=2\columnwidth]{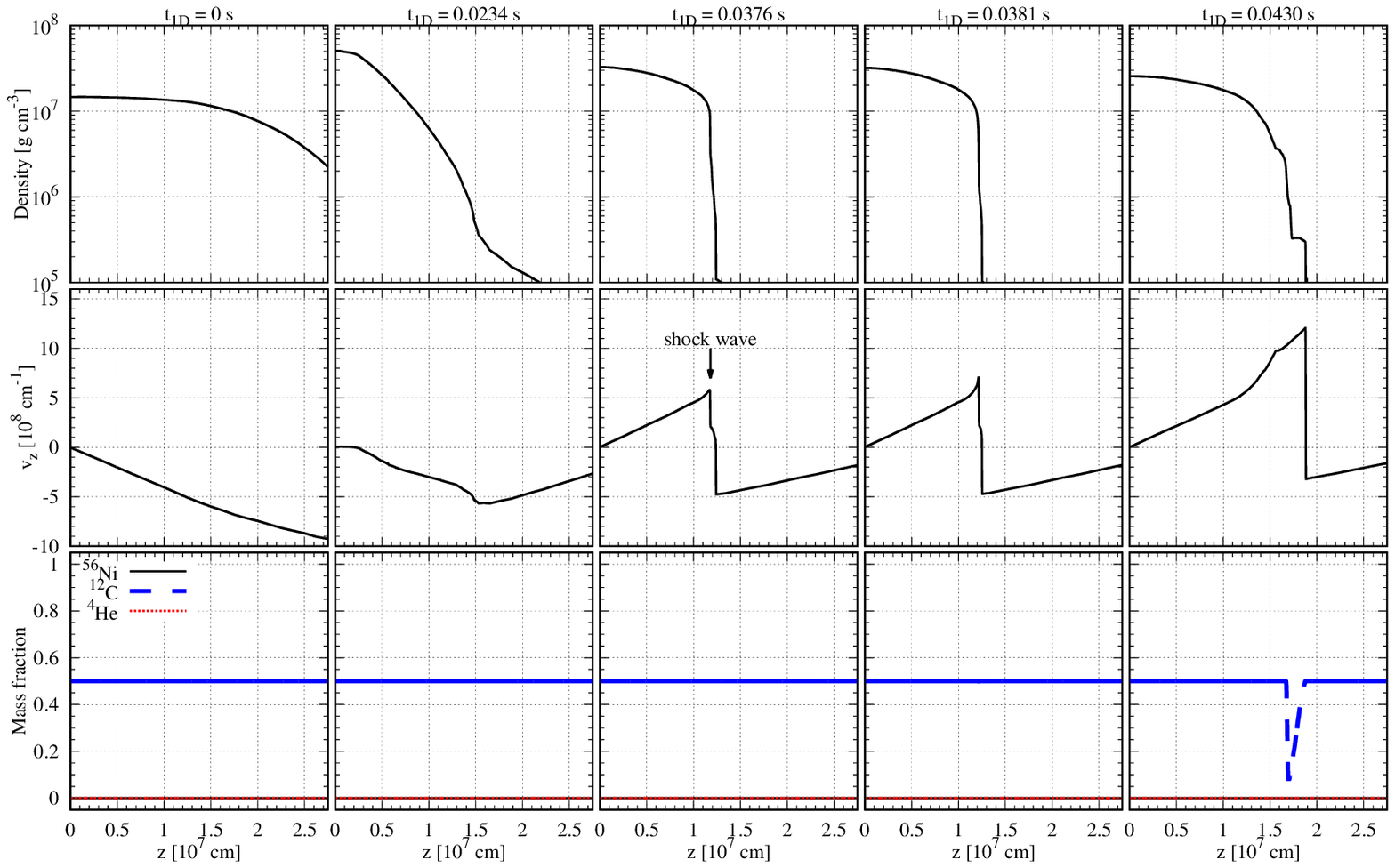}
  \caption{The same as Figure~\ref{fig:evolve1d_cohe}, except for a
    CO~WD without a He shell. \label{fig:evolve1d_cowd}}
\end{figure*}

Figure~\ref{fig:evolve1d_cowd} shows the time evolution in 1D mesh
simulation for a CO~WD without a He shell. The evolution of the WD is
the same as a CO~WD with a He shell from the beginning until the
emergence of a shock wave ($\tone = 0.0376$~s). However, the shock
wave excites nuclear reactions slightly, and consumes small amounts of
C materials (see the bottom panel at $\tone = 0.0430$~s). The nuclear
reactions cannot trigger further nuclear reactions. The nuclear
reactions cease soon. This CO~WD fails simple tidal detonation.

The shock wave successfully excites He detonation in the He shell of
the CO~WD in the following reason. The shock wave emerges in a region
with density of $\sim 10^7$~g~cm$^{-3}$. In front of the shock wave,
there is a region with density of $\sim 5 \times 10^6$~g~cm$^{-3}$ and
with size of $5 \times 10^5$~cm. This is consistent with the
conditions of He detonation obtained by
\cite{2013ApJ...771...14H}. Note that our 1D mesh simulation has
sufficiently high resolution ($\sim 10^4$~cm) to resolve the required
size ($\sim 10^5$~cm). As described above, the 1D mesh simulation
overestimates density by about $20$\% at the edge of the WD where the
shock wave and He detonation wave appear. It underestimates the size
of a hotspot required to generate a detonation wave by a factor of
$2$, since the size of a hotspot to generate a detonation wave
increases by three orders of magnitude with density decreasing by an
order of magnitude \citep{2013ApJ...771...14H}. The size of a region
heated by the shock wave is $5 \times 10^5$~cm, while the size of a
hotspot required to generate a detonation wave would be $< 5 \times
10^5$~cm. Eventually, the density overestimate in the 1D mesh
simulation does not affect the initiation of the He detonation wave.

On the other hand, the shock wave fails CO detonation in the CO~WD
without a He shell. The shock wave emerges in the same position as
that in the CO~WD with a He shell. However, the composition at the
position is CO composition, not He composition. According to
\cite{2009ApJ...696..515S}, CO detonation succeeds under density
environment of $\sim 5 \times 10^6$~g~cm$^{-3}$ only if a hotspot size
is $\gtrsim 10^6$~cm. However, the shock wave heats a region whose
size is several $10^5$~cm. Our results are consistent with
\cite{2009ApJ...696..515S}.

In the CO~WD with a He shell, the He detonation directly drives the CO
detonation. This is so-called ``edge-lit'' type of the double
detonation scenario in the context of type Ia supernovae. For the
edge-lit type, the altitude of the initiation point of He detonation
from the CO core-He shell interface should be $\gtrsim 10^7$~cm
\citep{2013ApJ...774..137M}. On the other hand, the edge-lit CO
detonation succeeds in our model despite that the altitude is several
$10^5$~cm (see the bottom panel at $\tone = 0.0376$~s). We can resolve
this discrepancy considering the motion of the CO core-He shell
interface. The interface in the case of \cite{2013ApJ...774..137M} is
at rest, while the interface in our model proceeds toward the He
detonation at a speed of $> 5 \times 10^8$~cm~s$^{-1}$ (see the middle
and bottom panels at $\tone = 0.0376$~s and $0.0381$~s). Therefore,
the edge-lit CO detonation in the CO core is easier to occur in our
model than in the case of \cite{2013ApJ...774..137M}.

\section{Conclusion}
\label{sec:conclusion}

We show tidal double detonation actually occurs in a WD~TDE by
numerical simulation. The WD is a CO~WD with a He shell whose mass
fraction is at most $5$\%. Importantly, if CO~WDs with and without a
He shell are in the same orbit around an IMBH, tidal double detonation
in the CO~WD with a He shell is easier to arise than tidal detonation
in the CO~WD without a He shell. In other words, tidal double
detonation emerges for smaller $\beta$ than tidal detonation. This
means tidal double detonation spreads opportunity to TDEs illuminating
IMBHs.

We may distinguish tidal double detonation from simple tidal
detonation by the presence of surface radioactivity from $^{56}$Ni
synthesized by He detonation. As seen in the bottom panel at $\tone =
0.0430$ in Figure~\ref{fig:evolve1d_cohe}, He detonation yields larger
mass fraction of $^{56}$Ni than CO detonation under similar density
environments ($\sim 10^7$~g~cm$^{-3}$). Therefore, WD~TDEs powered by
tidal double detonation have earlier emission than those powered by
simple tidal detonation, analogously to type Ia supernovae possibly
with He detonation
\citep[e.g.][]{2014Sci...345.1162D,2016MNRAS.459.4428K,2017Natur.550...80J}.

\section*{Acknowledgments}

A. Tanikawa thanks I. Hachisu for fruitful discussions. Numerical
computations were carried out on Cray XC30 at Center for Computational
Astrophysics, National Astronomical Observatory of Japan, on Cray XC40
at Yukawa Institute for Theoretical Physics, Kyoto University, and on
Oakforest-PACS at Joint Center for Advanced High Performance
Computing. The software used in this work was in part developed by the
DOE NNSA-ASC OASCR Flash Center at the University of Chicago. This
research has been supported in part by MEXT program for the
Development and Improvement for the Next Generation Ultra High-Speed
Computer System under its Subsidies for Operating the Specific
Advanced Large Research Facilities, and by Grants-in-Aid for
Scientific Research (16K17656, 17H06360) from the Japan Society for
the Promotion of Science.

\bsp	
\label{lastpage}
\end{document}